\documentclass[twocolumn,showpacs,preprintnumbers,amsmath,amssymb]{revtex4}

\input psfig.sty
\usepackage{graphicx}% Include figure files
\usepackage{dcolumn}% Align table columns on decimal point
\usepackage{bm}% bold math

\begin{document}

%\documentstyle[prd,epsfig,aps,preprint]{revtex}
%\documentstyle[prd,epsfig,aps]{revtex}
%\documentstyle[12pt,psfig]{article}
%\begin{document}

%\pretolerance{100}
%\draft
\title{Deflationary $\Lambda(t)$ cosmology: observational expressions}

\author{J. V. Cunha} \email{jvital@dfte.ufrn.br}
% \altaffiliation[Also at ]{Physics Department, XYZ University.}%Lines break automatically or can 
%be forced with \\
\author{J. A. S. Lima}
 \email{limajas@dfte.ufrn.br} 
\author{N. Pires}
 \email{npires@dfte.ufrn.br}%
\affiliation{%
Departamento de F\'{\i}sica, Universidade Federal do Rio Grande do Norte, C.P. 1641, 59072-970, 
Natal, RN, Brasil
}%

\date{\today}% It is always \today, today,

%\date{\today}
%\maketitle

%\vskip 1.5cm

\begin{abstract}
We discuss the classical cosmological tests for a large
class of FRW type models driven by a decaying vacuum energy
density. Analytic expressions for the lookback time, age of the
universe, luminosity distance, angular diameter, and galaxy number
counts versus redshift are derived and their meaning discussed in
detail. It is found that the standard FRW results are
significantly altered, showing that such tests may constrain the
physical parameters of these models which are also in agreement to
the accelerated expansion suggested by the latest SNe type Ia
observations.
\end{abstract}


\maketitle

%\newpage
\section{Introduction}

Recent measurements from some type Ia Supernovae (SNe) at
intermediate and high redshifts (Perlmutter et al. 1999, Riess et
al. 1998) indicate that the bulk of energy in the Universe is
repulsive and appears like a ``quintessence" component, that is,
an unknown form of dark energy (in addition to the ordinary CDM
matter) probably of primordial origin (see Turner 2000 for a
review). Together with the observations of CMB anisotropies (de
Bernardis 2000), such results seem to provide an important piece
of information connecting an early inflationary stage with the
astronomical observations.

This state of affairs has stimulated the interest for more general
models containing an extra component describing this dark energy,
and simultaneously accounting for the present accelerated stage of
the Universe. However, the absence of a convincing evidence on the
nature of the dark component gave origin to an intense debate and
mainly to theoretical speculations. A possible list of old and new
candidates for ``quintessence" now include:

(i) a decaying vacuum energy density, or a time varying
$\Lambda$-term (\"{O}zer and Taha 1987, Freese at al. 1987, Chen
and Wu 1990, Carvalho et al. 1992, Waga 1993; for reviews see
Overduin and Cooperstoock 1998; Sahni and Starobinski 2000)

(ii) the so-called ``X-matter", an extra component simply
characterized by an equation of state $p_x=\omega\rho_{x}$, where
$\omega\geq -1$ (Turner and White 1997, Chiba et al. 1997,
Efstathiou 1999, Lima and Alcaniz 2000, Turner 2000, Alcaniz and
Lima 2001), which describes, as a particular case, cosmologies
with a constant $\Lambda$-term ($\Lambda$CDM models). Generically,
the $\omega$ parameter may be a function of the redshift (Cooray
and Huterer 2000)

(iii) a rolling scalar field (Ratra and Peebles 1988, Caldwell et
al. 1998, Wang et al. 2000).

Here we are interested in the first class of models. The basic
reason is the widespread belief that the early Universe evolved
through a cascade of phase transitions, thereby yielding a vacuum
energy density which at present is at least $118$ orders of
magnitude smaller than in the Planck time (Weinberg 1989). Such a
discrepancy between theoretical expectation (from the modern
microscopic theory of particles and gravity) and empirical
observations constitutes a fundamental problem in the interface
uniting astrophysics, particle physics and cosmology, which is
often called ``the cosmological constant problem'' (Weinberg 1989;
Jack NG 1992; Dolgov 1997). This puzzle inspired some authors
(Lima and Maia 1994, Lima and Trodden 1996) to propose a class of
phenomenological deflationary cosmologies driven by a decaying
vacuum energy density where the present value, $\Lambda_o =
\Lambda (t_o)$, is a remnant of the primordial inflationary stage
(from now on the subscript ``o" denotes the present day
quantities). The basic scenario has an interesting cosmological
history that evolves in three stages. Initially, an unstable de
Sitter configuration, with no matter and radiation is supported by
the largest values of the vacuum energy density. This nonsingular
de Sitter state evolves to a quasi-FRW vacuum-radiation-dominated
phase, and, subsequently, the Universe changes continuously from
vacuum-radiation to the present vacuum-dust dominated phase. The
first stage harmonizes the scenario with the cosmological constant
problem, while the transition to the second stage solves the
horizon and other well-know problems in the same manner as in
inflation. Finally, the Universe enters in the present vacuum-dust
phase with a negative deceleration parameter as required by the
SNe type Ia observations.

In this article, we focus our attention on this class of
deflationary decaying vacuum models. The effective time dependent
cosmological term is regarded as a second fluid component with
energy density, $\rho_v(t) =\Lambda(t)/{8 \pi G}$, which transfers
energy continuously to the material component. The main goal is to
investigate the basic kinematic tests in the present vacuum-dust
dominated phase, or equivalently, how the classical cosmological
tests may constrain the physical parameters of such models. The
paper is organized as follows: In section 2, we set up the basic
equations for deflationary cosmologies driven by a decaying
$\Lambda(t)$-term. In section 3, the expressions for classical
cosmological tests are derived and compared with the conventional
expressions without the $\Lambda$-term. Section 4 gives the
conclusion of the main results, and, in the appendix A, the exact
expression yielding the dimensionless radial coordinate as a
function of the redshift is deduced.

\section{Deflationary $\Lambda(t)$ Cosmology : Basic Equations}

We shall consider a class of spacetimes described by the general
FRW line element ($c=1$)
\begin{equation}
 ds^2 = dt^2 - R^{2}(t) \left(\frac{dr^2}{1-k r^2} + r^2 d \theta^2
+ r^2 sin^{2}\theta d \phi^2\right) ,
\end{equation}
where $R(t)$ is the scale factor, $k=0$, $\pm 1$ is the curvature
parameter of the spatial sections, and $r$, $\theta$ and $\phi$
are dimensionless comoving coordinates. In that background, the
Einstein field equations (EFE) with a nonvacuum component plus a
cosmological $\Lambda(t)$-term are:
\begin{equation}
\label{rho} 8\pi G \rho + \Lambda(t) = 3 \frac{\dot{R}^2}{R^2} + 3
\frac{k}{R^2}\quad,
\end{equation}
\begin{equation}
\label{press} 8\pi G p - \Lambda(t) = -2 \frac{\ddot{R}}{R} -
\frac{\dot{R}^2}{R^2} - \frac{k}{R^2}\quad ,
\end{equation}
where an overdot means time derivative, $\rho$ and $p$ are the
energy density and pressure, respectively. As usual, we consider
that the nonvacuum component obeys the ``$\gamma$-law" equation of
state
\begin{equation}
\label{lgamma} p = (\gamma -1)\rho \quad ,
\end{equation}
where $\gamma\in[1,2]$ specifies if the fluid component is
radiation ($\gamma={4 \over 3}$) or dust ($\gamma=1$).

Phenomenologically, we also assume that the effective
$\Lambda(t)$-term is a variable dynamic degree of freedom so that
in an expanding universe it relaxes to its present value according
with the following ansatz (Lima and Trodden 1996)
\begin{equation}
\label{ansatz} \rho_{v} = \frac{\Lambda(t)}{8 \pi G} = \beta
\rho_{T} \left(1 + \frac{1 - \beta}{\beta} {H \over H_{I}}\right)
\quad ,
\end{equation}
where $\rho_{v}$ is the vacuum density, $\rho_{T}=\rho_{v}+ \rho$
is the total energy density, $H={\dot{R}}/R$ is the Hubble
parameter, $H_{I}^{-1}$ is the arbitrary time scale characterizing
the deflationary period, and $\beta\in[0,1]$ is a dimensioneless
parameter of order unity.

It is worth noticing that for $H=H_{I}$, the above definition
reduces to $\rho_{v}=\rho_{T}$ so that we have inflation with no
matter-radiation component ($\rho=0$). In particular, if
$H_{I}^{-1}$ is the order of the Planck time, that is,
$H_{I}^{-1}\sim 10^ {-43}$s, one may show that such models are in
accordance with the cosmological constant problem in the sense
that ${\Lambda_{Planck}/\Lambda_o}\sim 10^{118}$.

At late times ($H << H_{I}$), we see from (5) that $\rho_{v} \sim
\beta \rho_{T}$, as required by the recent Supernovae
observations. To be more precise, if the deflationary process
begins at Planck time, the ratio $H_{0}/H_{I} \sim 10^{-60}$ while
the remaining terms are of order unity. Indeed, even if deflation
begins much later, say at $10^{-35}$s, or even at $10^{-15}$s (the
respective scales of grand and electroweak unification in the
standard model), one obtains $H_{0}/H_{I} \sim 10^{-52}$ and
$H_{0}/H_{I} \sim 10^{-32}$, respectively. This means that to a
high degree of accuracy, the scale $H_{I}$ is unimportant during
the vacuum-dust dominated phase. Therefore, since in this work we
mainly interested in the classical cosmological tests, henceforth
we consider only this limiting behavior with the material
component described by a pressureless fluid ($\gamma=1$).

Let us now consider the evolution of the scale factor. Combining
equations (\ref{rho})-(\ref{ansatz}) and taking the limit
$H<<H_I$, it is readily seen that the scale factor during the
vacuum-dust phase obeys the slightly modified FRW equation

\begin{equation}\label{ddR}
2R\ddot{R} +  (1-3\beta)\dot{R}^2 +  (1-3\beta)k = 0 \quad ,
\end{equation}
the first integral of which is
\begin{equation}
\label{dR} {\dot{R}}^2 =  \frac{A}{R^{1-3\beta}} - k \quad ,
\end{equation}
where the constant $A>0$ in order that $\rho$ be positive definite
in this phase (see Eq.(2)). By expressing the constant $A$ in
terms of the present day parameters, it is straightforward to show
that the above equation can be rewritten as
\begin{equation}
\label{dR1} \left({\dot{R} \over R_{o}}\right)^{2} = H_{o}^{2}
\left[1 - \left(\frac{\Omega_{o}} {1 - \beta}\right) +
\left(\frac{\Omega_{o}}{1 - \beta}\right)\left({R_{o}
 \over R}\right)^{1-3\beta} \right]\quad .
\end{equation}
where $\Omega_o \equiv {\rho \over \rho_c}\vert_{t=t_o}$ is the
present value of the matter density parameter. For $\beta=0$ the
above equation reproduces the standard  cold dark matter FRW
result (Kolb and Turner 1990). Whereas for $\beta\neq 0$, it
describes the influence of the decaying vacuum in the present
phase.

Following standard lines we also define the deceleration parameter
$q_o = - {{R\ddot R \over {\dot R}^2}}\vert_{t=t_o}$. Using
equations (\ref{ddR}) and  (\ref{dR}) one may show that
\begin{equation}
q_o = {{1-3\beta}\over 2} \left(\frac{\Omega_o}{1 - \beta}\right)
\quad.
\end{equation}
Then, for any value of $\Omega_o \neq 0$, we see that the
deceleration parameter $q_o$ with decaying vacuum energy is always
smaller than the corresponding one of the FRW model. The critical
case ($\beta={1 \over 3},q_o = 0$), describes a ``coasting
cosmology". However, instead of being supported by ``K-matter"
(Kolb 1989), this kind of model is obtained in the present context
for a vacuum-dust filled universe, and the corresponding solutions
hold regardless of the value of $\Omega_o$. It is also interesting
that even negative values of $q_o$ are allowed since the
constraint $q_o < 0$ can always be satisfied provided $\beta >
1/3$. These results are in line with recent measurements of the
deceleration parameter $q_o$ using Type Ia supernovae (Perlmutter
et al. 1999; Riess et al. 1998). Such observations indicate that
the universe may be accelerating today, which corresponds
dynamically to a negative pressure term in the EFE. For a fixed
$\Omega_o$, this means that the universe with decaying vacuum is
older than the corresponding FRW model with the usual deceleration
parameter $q_o \geq 0$. This behavior also reconcile other recent
results (Freedman 1998), pointing to a Hubble parameter $H_o$
larger than $50$ km s$^{-1}$ Mpc$^{-1}$.

Now, combining equations (\ref{rho}), (\ref{ansatz}) and
(\ref{dR}) for ($H\ll H_{I}$) one finds that vacuum and the matter
energy density can be expressed as
\begin{equation}
\label{rhovac} \rho_{v} = \beta \rho_{T} = \beta
\rho_{T0}\left(\frac{R_{0}} {R}\right)^{3(1 - \beta)}\quad ,
\end{equation}
\begin{equation}
\label{rhomat} \rho = (1 - \beta) \rho_{T} = (1 - \beta)\rho_{T0}
\left(\frac{R_{0}}{R}\right)^{3(1 - \beta)}\quad ,
\end{equation}
where $\rho_{T0} = 3A/8 \pi G R_{0}^{3(1 - \beta)}$. In virtue of
the decaying vacuum energy, we also see that the explicit
dependence of the energy density on the scale factor $R(t)$ is
slightly modified in comparison with the standard case. As should
be expected from Eqs. (\ref{ddR}) and (\ref{dR}), some dynamic
expressions may be obtained from the standard FRW ones simply by
replacing the ``index" $\gamma =1$ by an effective parameter
$\gamma_{eff} = 1 - \beta$ (see the above scale laws).

\section{Kinematic Tests}

The kinematical relation distances must be confronted with the
observations in order to put limits on the free parameters of the
deflationary class of models presented in the last section. As
remarked before, $H_{I}$ cannot be constrained by the classical
tests since it is unimportant in the present phase. It is adjusted
in order to put the models in accordance to the cosmological
constant problem (Lima and Trodden 1996). It thus follows that we
have only the pair of free parameters ($\Omega_o, \beta$)
satisfying $\Omega_{T0}=\Omega_{0}/(1-\beta)$, where $\Omega_{T0}$
is the present value of the total density parameter.

\begin{figure}
\vspace{.2in}
\centerline{\psfig{figure=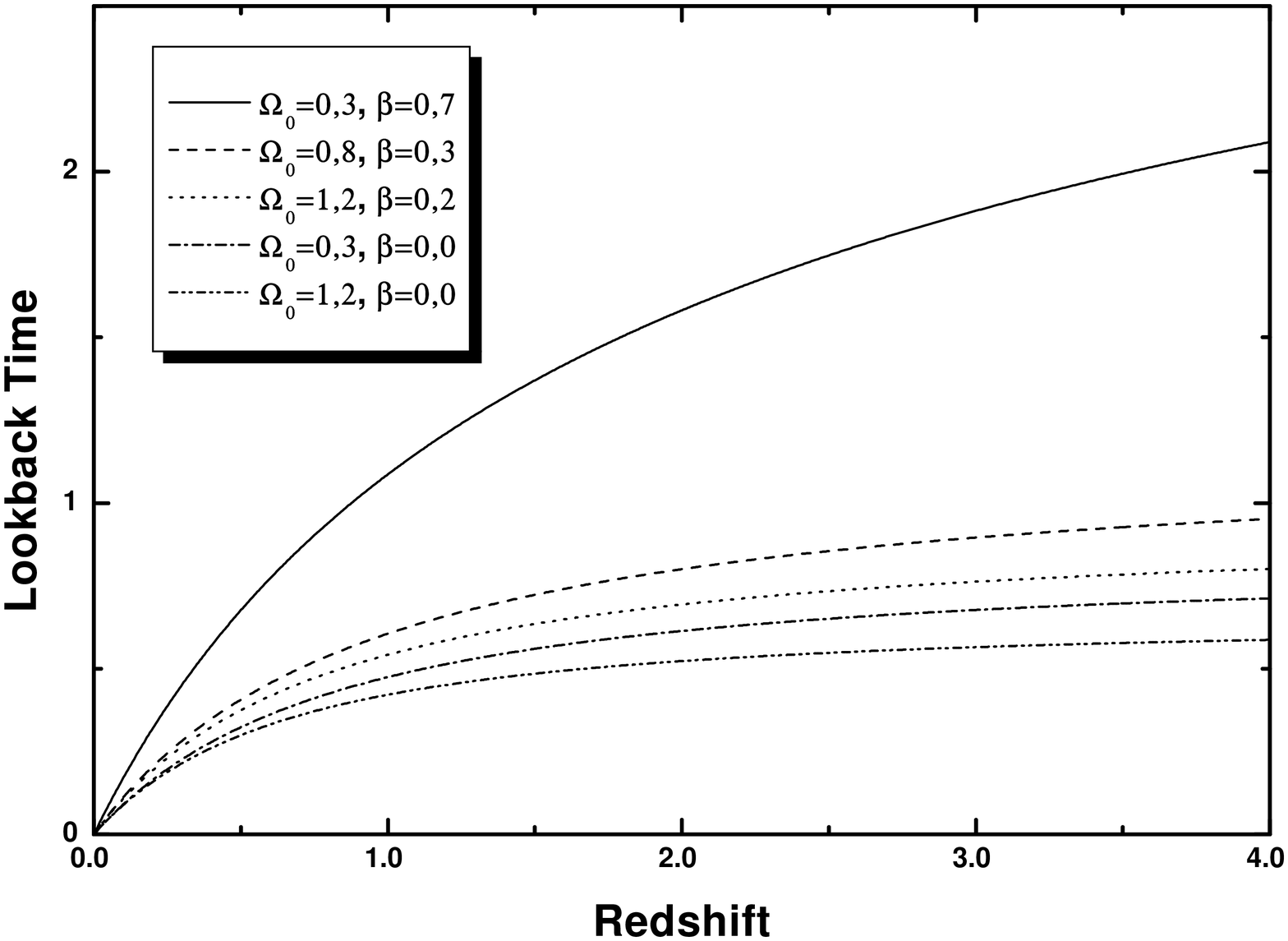,width=8.5truecm,height=7truecm}
\hskip 0.1in} \caption{Lookback time as a function of the redshift
for some selected values of $\Omega_{0}$ and $\beta$. As should be
expected, the difference $t_o - t(z)$ increases with the vacuum
energy contribution (larger values of $\beta$).} 
\end{figure}

a) {\it Lookback time-redshift diagram}

The lookback time, $\Delta t = t_o - t(z)$, is the difference
between the age of the universe at the present time ($z=0$) and
the age of the universe when a particular light ray at redshift
$z$ was emitted. By integrating (\ref{dR1}) such a quantity is
easily derived
\begin{equation}
t_o - t(z) = {H_o}^{-1} \int\limits_{1 \over 1+z}^{1} {dx \over
\sqrt{1 - \left(\frac{\Omega_o}{1 - \beta}\right) +
\left(\frac{\Omega_o}{1 - \beta}\right) x^{-(1-3\beta)}}},
\end{equation}
which generalizes the standard FRW result (Kolb and Turner 1990).
The age of the universe is obtained by taking the limit $z
\rightarrow \infty$  in  the above equation. We find

\begin{equation}
t_o = {H_o}^{-1} \int_{0}^{1} {dx \over \sqrt{1 -
\left(\frac{\Omega_o}{1 - \beta}\right) + \left(\frac{\Omega_o}{1
- \beta}\right) x^{-(1-3\beta)}}} \quad .
\end{equation}

For $\beta=0$ the above expressions reduce to the ones of the
standard FRW models (Kolb and Turner 1990). Generically, we see
that decaying vacuum increases the dimensionless parameter
$H_{o}t_{o}$ while preserving the overall expanding FRW behavior.
The lookback time curves as a function of the redshift for some
selected values of $\Omega_{o}$ and $\beta$ are displayed in Fig.
1. For completeness, in Fig. 2 we show the age of the Universe (in
units of $H_{o}$) as a function of $q_o$ for some selected values
of $\beta$.

\begin{figure}
\vspace{.2in}
\centerline{\psfig{figure=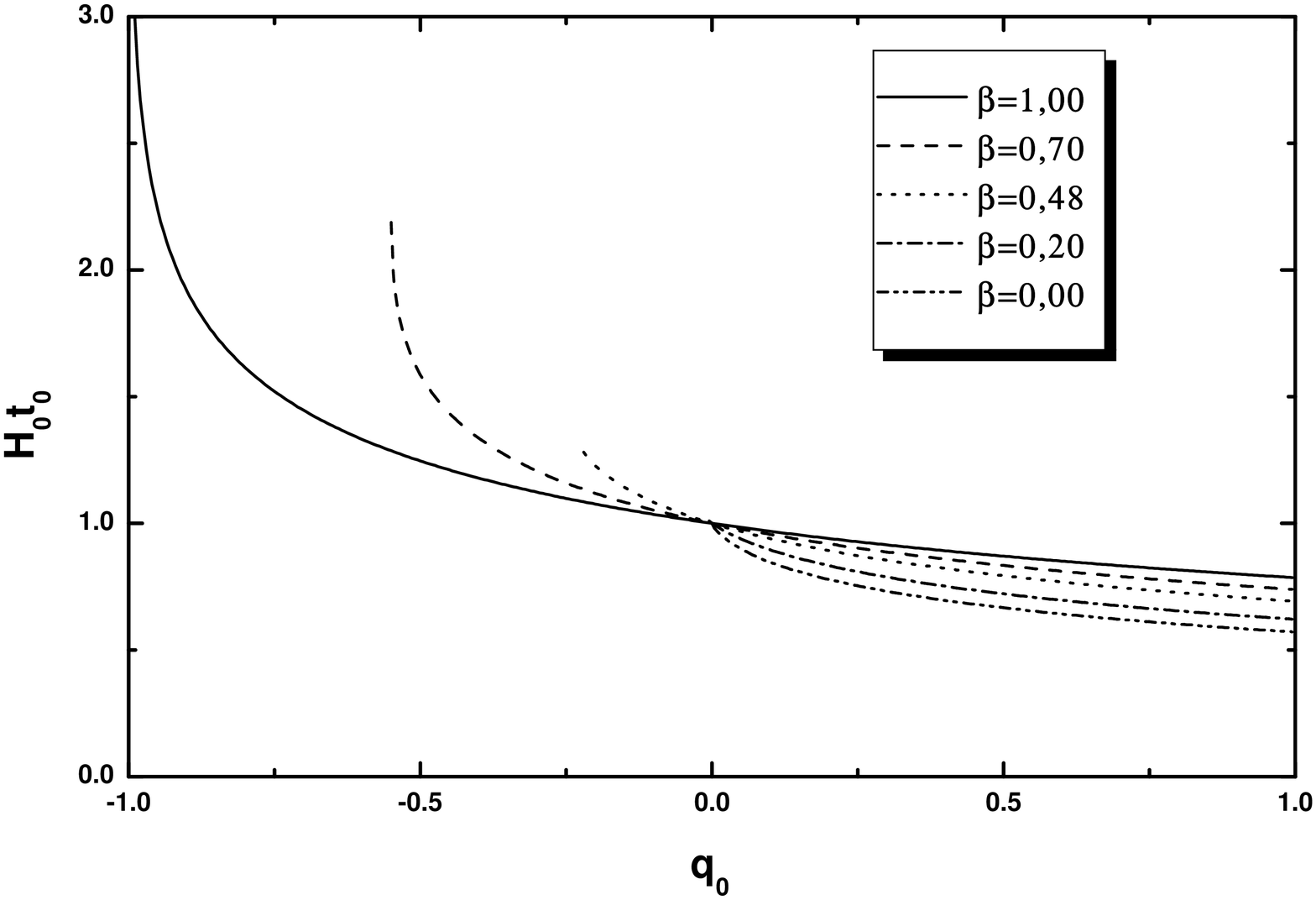,width=8.5truecm,height=7truecm}
\hskip 0.1in} \caption{Age parameter as a function of the
deceleration parameter for some selected values of $\beta$.}
\end{figure}

b) {\it Luminosity distance-redshift}

The luminosity distance of a light source is defined as the ratio
of the detected energy flux $L$, and the apparent luminosity,
i.e., $d_L^{2} = {L \over 4 \pi l}$. In the homogeneous and
isotropic FRW metric (1) it takes the form below (Sandage 1988)
\begin{equation}
d_L = R_o r_1(z)(1 + z)\quad ,
\end{equation}
where $r_1(z)$ is the radial coordinate distance of the object at
light emission. Inserting $r_1(z)$ derived in the Appendix, it
follows that
\begin{equation}
 d_L = \frac{(1+z)sin[\delta sin^{-1}(\alpha_1)-\delta
sin^{-1}(\alpha_2)]}{H_o \left(\frac{\Omega_o}{1-\beta} - 1
\right)^{\frac{1}{2}}}\quad .
\end{equation}
where $\delta=\frac{2}{(1-3\beta)}$, $\alpha_1=
(1-\frac{1-\beta}{\Omega_o})^{{1 \over 2}}$ and
$\alpha_{2}=\alpha_{1} (1 + z)^{-(\frac{1-3\beta}{2})}$.

\begin{figure}
\vspace{.2in}
\centerline{\psfig{figure=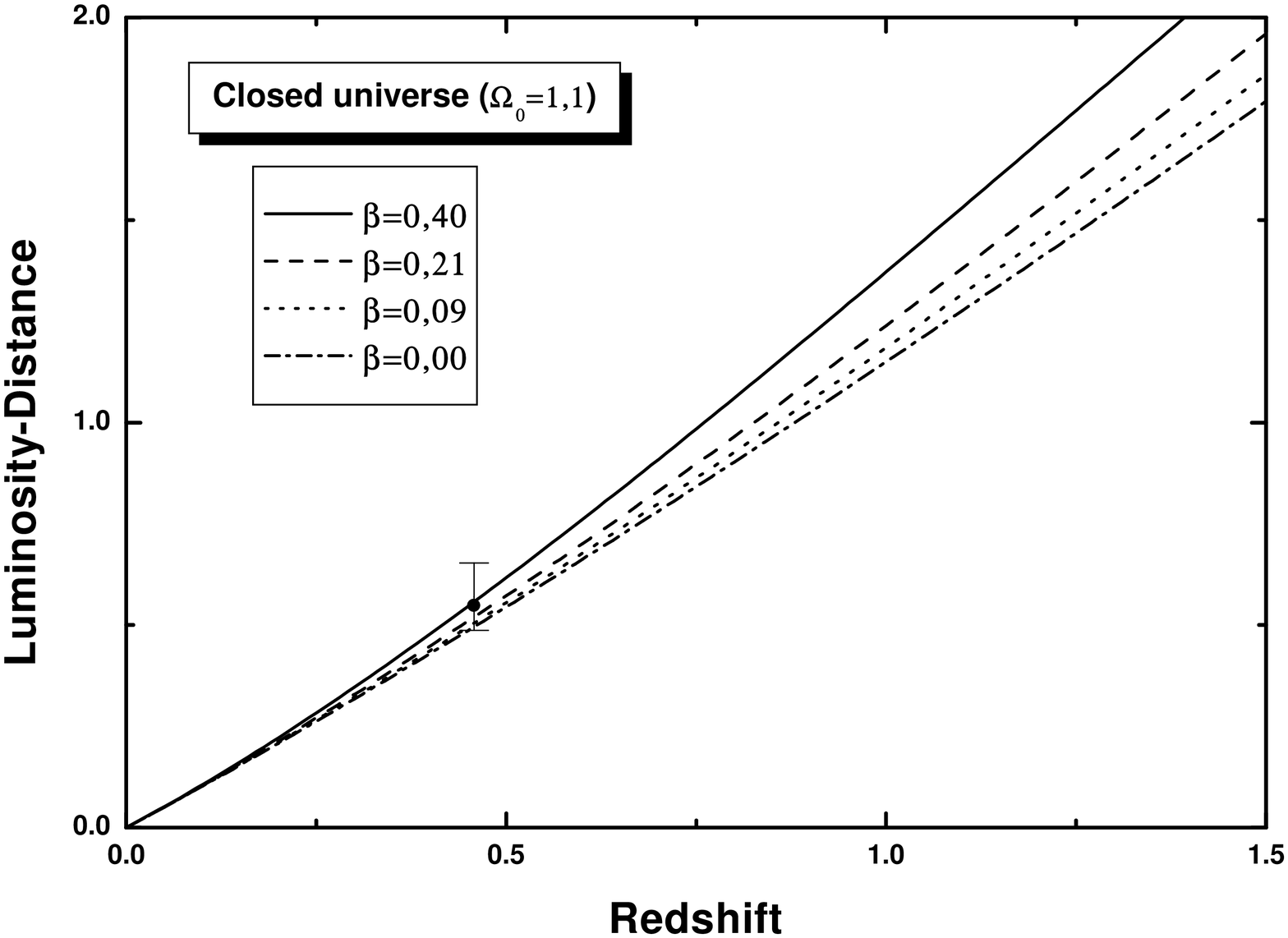,width=8.5truecm,height=7truecm}
\hskip 0.1in} \caption{Luminosity distance as a function of the
redshift for closed models with decaying vacuum energy. Here and
in Fig. 4 the typical error bar and data point are taken from
Perlmutter et al. (1995)}
\end{figure}

As one may check, expressing $\Omega_o$ in terms of $q_o$ from
(9), and taking the limit $\beta \rightarrow0$, the above
expression reduces to
\begin{equation}
 d_L = \frac{1}{{H_o}q_o^{2}}[zq_o + (q_o - 1)(\sqrt{2q_oz + 1}
-
1)]\quad ,
\end{equation}
which is the usual FRW result (Weinberg 1972). Expanding eq.(15)
for small $z$ gives
\begin{equation}
H_o d_L = z + \frac{1}{2}\left[1 - \frac{{1 -
3\beta}}{2}\left(\frac{\Omega_o}{1 - \beta}\right)\right] z^{2}
+...\quad ,
\end{equation}
which depends explicitly on the $\beta$ parameter. However,
replacing $\Omega_o$ from (9) we recover the usual FRW expansion
for small redshifts, which depends only on the effective
deceleration parameter $q_o$ (Weinberg 1972). This is not a
surprising result since expanding $d_L(z)$ in terms of $\Omega_o$,
the ``00'' component of Einstein's equations has implicitly been
considered, while the expansion in terms of $q_o$ comes only from
the form of the FRW line element. The luminosity distance as a
function of the redshift for closed and open models with decaying
vacuum energy is shown in Figures 3 and 4, respectively. As
expected for all kinematic tests, different cosmological models
have similar behavior at $z << 1$, and the greatest discrimination
among them comes from observations at large redshifts. A more
quantitive analysis based on the SNe observations will be
presented elsewhere.

\begin{figure}
\vspace{.2in}
\centerline{\psfig{figure=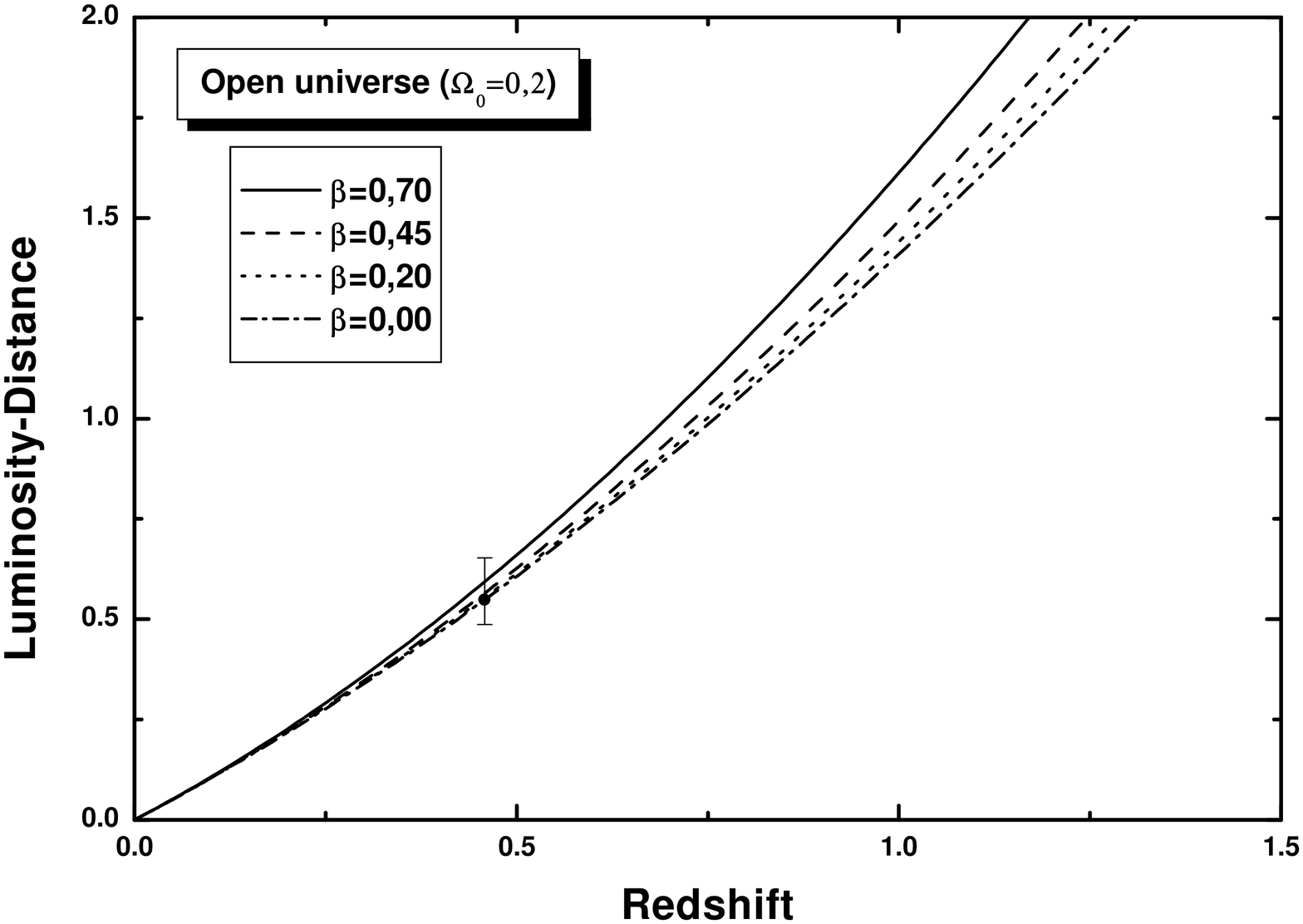,width=8.5truecm,height=7truecm}
\hskip 0.1in} \caption{Luminosity distance versus redshift. Solid
curve is the prediction for the standard FRW open universe
($\beta=0$).} 
\end{figure}

c) {\it Angular size-redshift}

Another important kinematic test is the angular size - redshift
relation $\theta(z)$. As widely known, the data concerning the
angular-size are until nowadays somewhat controversial (see
Buchalter et al. 1998 and references therein). Here we are
interested in angular diameters of light sources described as
rigid rods and not as isophotal diameters. These quantities are
naturally different, because in an expanding world the surface
brightness varies with the distance (Sandage 1988). The angular
size of a light source of proper size $D$ (assumed free of
evolutionary effects) located at $r = r_1(z)$ and observed at $r =
0$ is
\begin{equation}
\theta = \frac{D(1 + z)}{R_o r_1(z)}\quad .
\end{equation}
Inserting the expression of $r_1(z)$ given in the Appendix A it
follows that {\small{
\begin{equation}
\theta = \frac{DH_o \left(\frac{\Omega_o}{1 - \beta} - 1
\right)^{\frac {1}{2}}(1 + z)}{{sin[\delta sin^{-1}\alpha_1-\delta
sin^{-1}\alpha_2]}}\quad .
\end{equation}
}}

\begin{figure}
\vspace{.2in}
\centerline{\psfig{figure=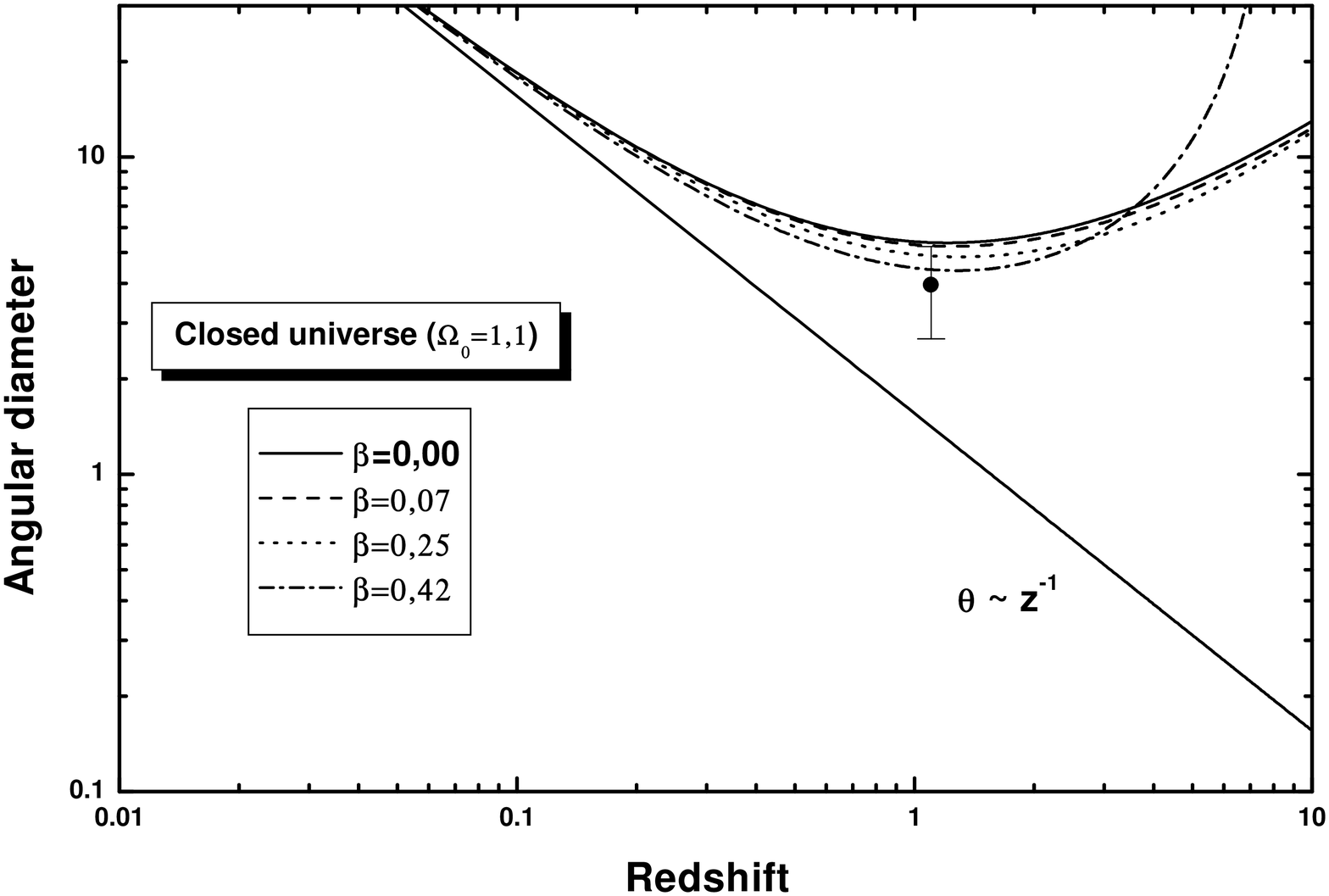,width=8.5truecm,height=7truecm}
\hskip 0.1in} \caption{Angular diameter versus redshift for closed
models with decaying vacuum energy and some selected values of
$\beta$. For comparison we also show the Euclidian result. Here
and in Fig 6, the typical error bar and data point are taken from
Gurvits et al. (1999).}
\end{figure}

For small $z$ one finds
\begin{equation}
\theta = {D H_o \over z}\left[ 1 + \frac{1}{2} \left(3 + \frac{{1
- 3\beta}}{2}(\frac{\Omega_o}{1 - \beta})\right) z
+...\right]\quad .
\end{equation}

Hence, decaying vacuum cosmologies as modeled here also requires
an angular size decreasing as the inverse of the redshift for
small $z$. However, for a given value of $\Omega_o$, the second
order term is a function only of the $\beta$ parameter. In terms
of $q_o$, inserting (9) into (20) it is readily obtained
\begin{equation}
\theta = \frac{D H_o}{z}[ 1 + \frac{1}{2} (3 + q_o) z +...]\quad ,
\end{equation}
which is formally the same FRW result for small redshifts (Sandage
1988). At this limit only the effective deceleration parameter may
be constrained from the data, or equivalently, at small redshifts
one cannot extract the values of $\Omega_o$ and $\beta$
separately. The angular size-redshift diagram for closed and open
models and selected values of the $\beta$ parameter is displayed
in Figures 5 and 6, respectively.

\begin{figure}
\vspace{.2in}
\centerline{\psfig{figure=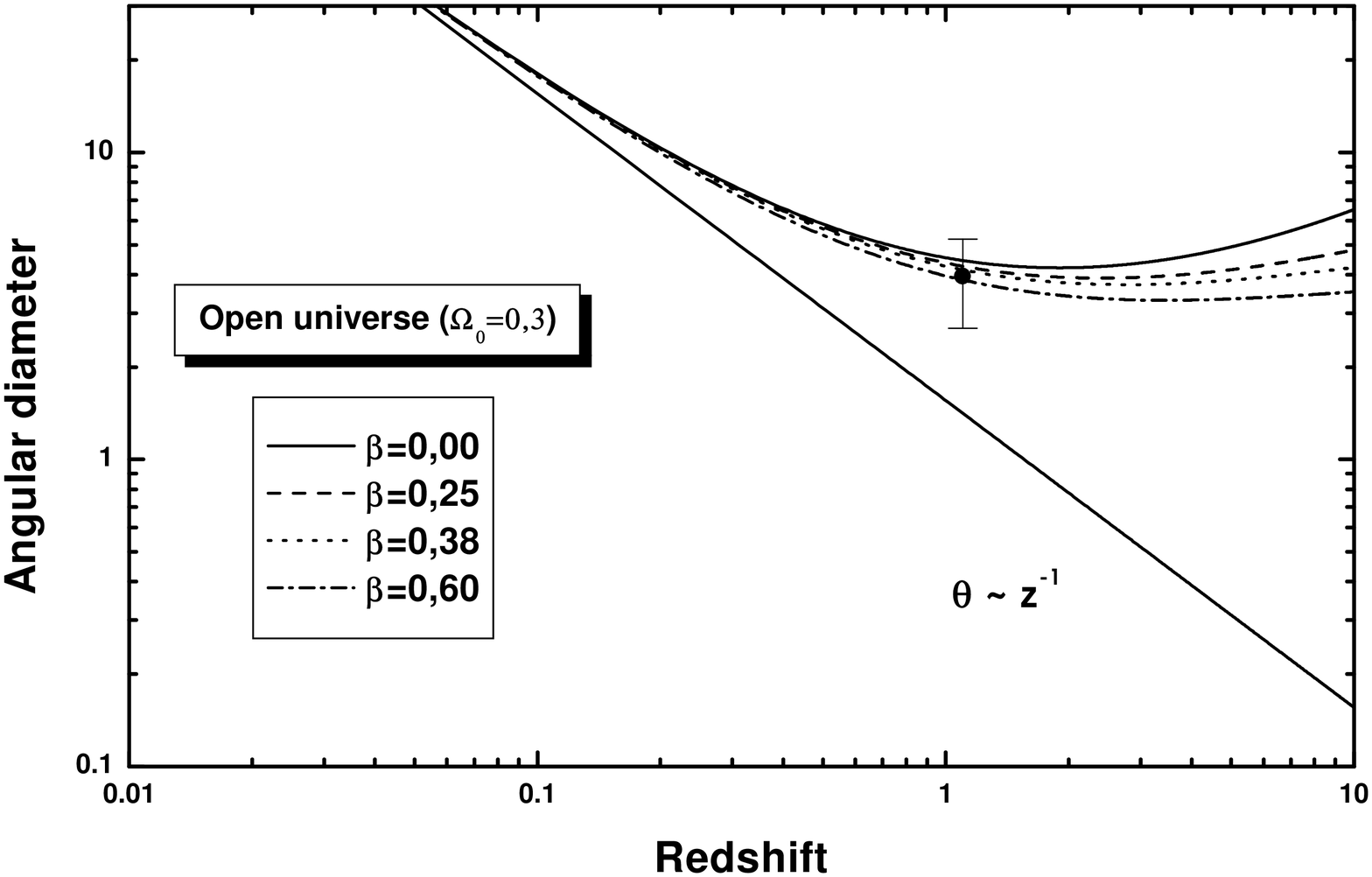,width=8.5truecm,height=7truecm}
\hskip 0.1in} \caption{The same graph of Fig. 5 for open models.}
\end{figure}

d) {\it Number counts-redshift}

The final kinematic test considered here is the galaxy number
count per redshift interval. The number of galaxies in a comoving
volume is equal to the number density of galaxies (per comoving
volume) $n_{g}$, times the comoving volume element $dV_{c}$
\begin{equation}
dN_{g}(z) = n_{g} dV_{c} = \frac{n_{g} r^{2} dr d{\Omega}
}{\sqrt{1 - k{r}^{2}}}\quad .
\end{equation}

\begin{figure}
\vspace{.2in}
\centerline{\psfig{figure=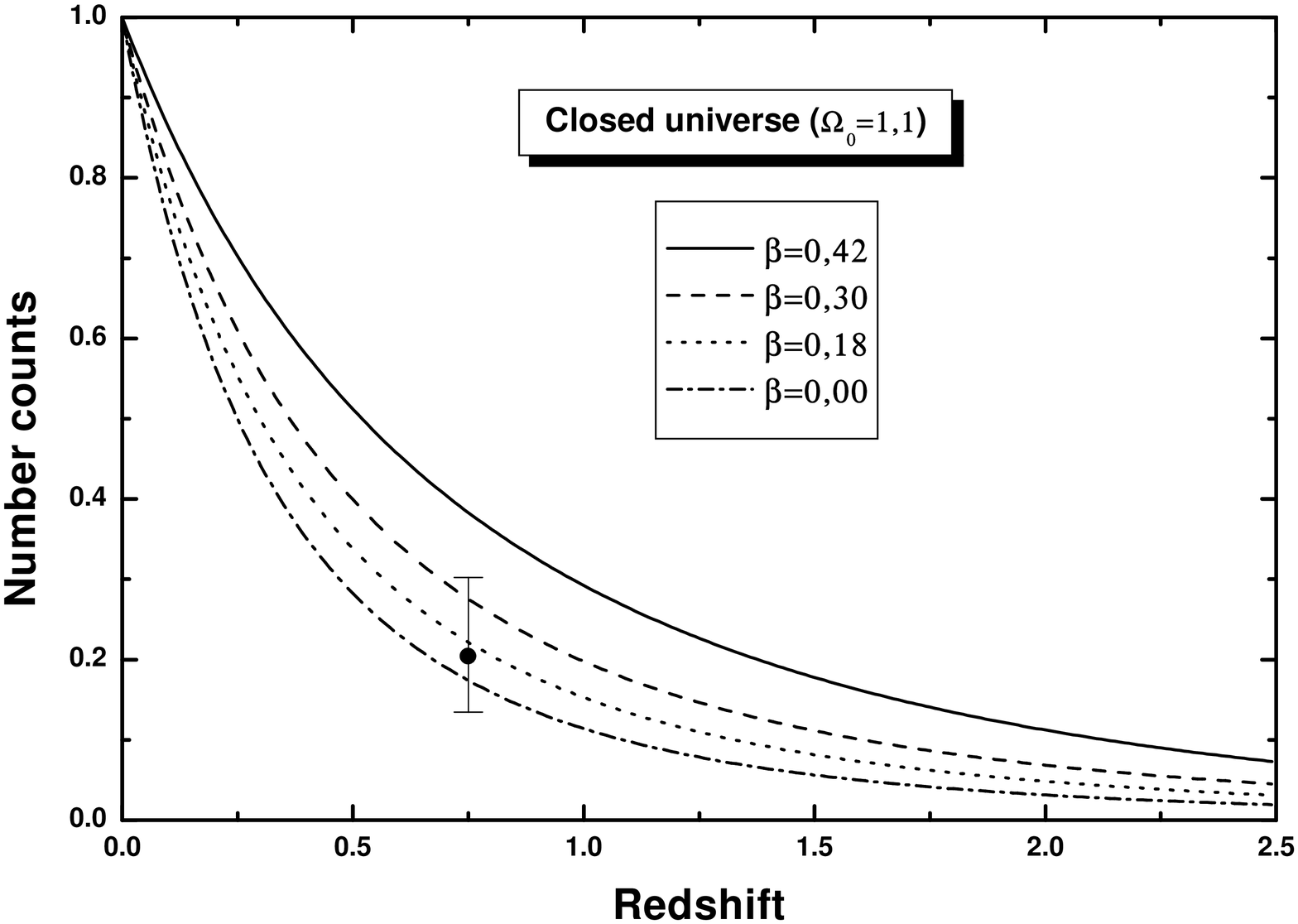,width=8.5truecm,height=7truecm}
\hskip 0.1in} \caption{Number counts as a function of the redshift
for deflationary closed models. Here and in Fig. 8 the typical
error bar and data point are taken from Loh \& Spillar (1986).}
\end{figure}

By using $r_1(z)$ as derived in appendix, it follows that the
general expression for number-counts can be written as
\begin{equation}
{(H_oR_o)^{3}dN_{g}\over n_{g}z^{2}dzd{\Omega}} =
{sin^{2}{\delta[sin^{-1} (\alpha_1)-sin^{-1}(\alpha_2)]} \over (1
+ z)z^{2}f(\Omega_o, \beta, z)}\quad ,
\end{equation}
where $f(\Omega_o, \beta, z) =
\left(\frac{\Omega_o}{1-\beta}-1\right)
\left[1-\frac{\Omega_o}{1-\beta}+\frac{\Omega_o}{1-\beta}(1+z)^{1-3\beta}\right]^{1/2}$. For small 
redshifts
\begin{equation}
\frac{(H_oR_o)^{3} dN_{g}}{n_{g}z^{2}dzd{\Omega}} = 1 -
2\left[\frac{(\frac{\Omega_o}{1 - \beta})(1-3\beta)}{2} +
1\right]z + ...\quad .
\end{equation}
In Figures 7 and 8 we have displayed the number counts-redshift
relations for flat, closed and open Universes for some selected
values of $\Omega_{o}$ and $\beta$. It is worth mentioning the
tendency of decaying vacuum models to have larger volumes per
redshift interval than the standard FRW models with the same
$\Omega_{o}$. This feature is similar to the one found in
``quintessence'' and $\Lambda$CDM cosmologies and could be
advantageous if the observational data indicate an excess count of
high-redshift objects. The limits on the $\beta$ parameter
obtained from all kinematic tests are shown in Table 1.

\begin{figure}
\vspace{.2in}
\centerline{\psfig{figure=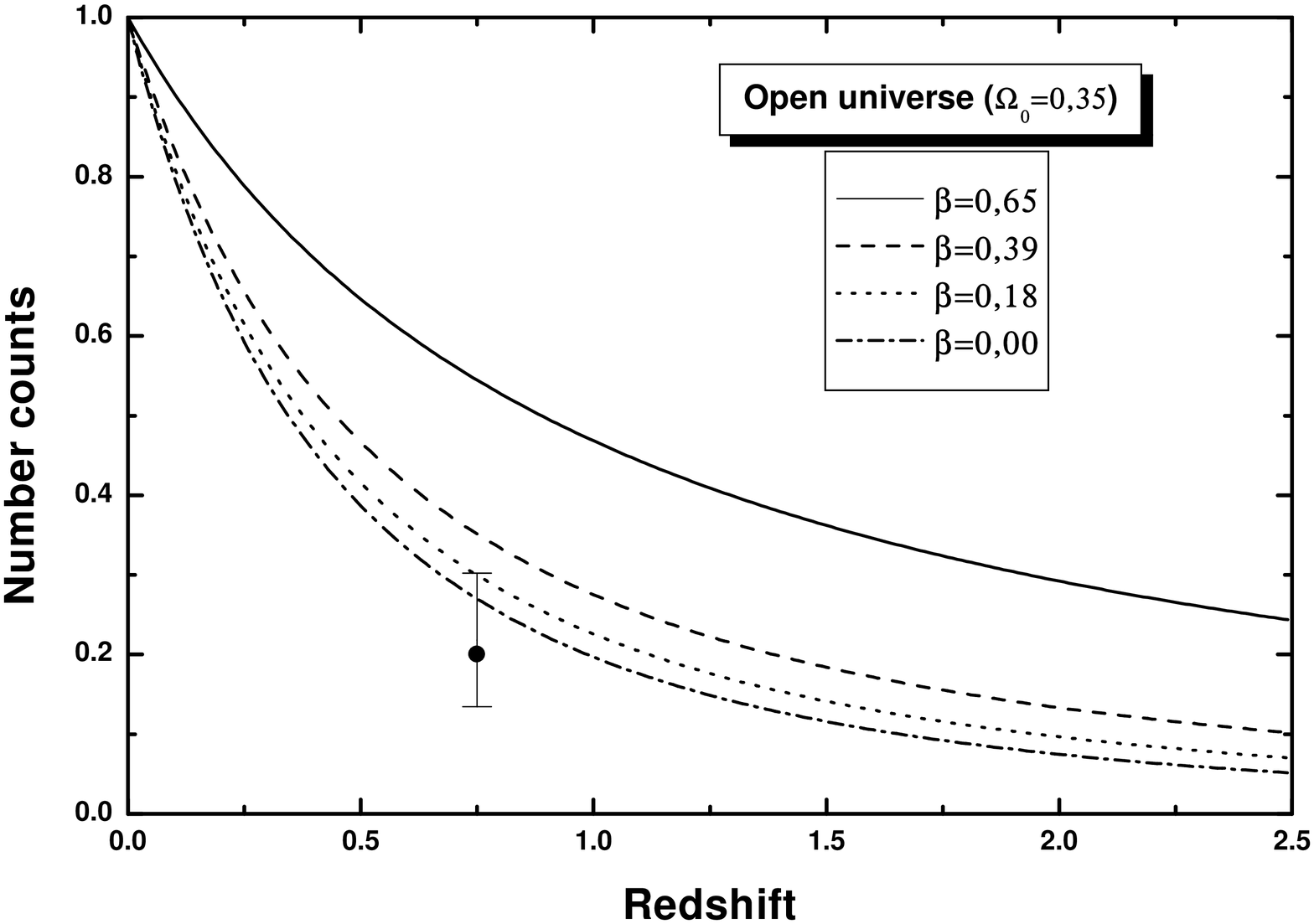,width=8.5truecm,height=6.5truecm}
\hskip 0.1in} \caption{Number counts versus redshift. The open
case.}
\end{figure}

At this point, it is also interesting to compare our results with
previous studies of kinematic tests for $\Lambda(t)$ cosmologies.
Observable expressions for a large class of flat decaying vacuum
models were discussed by Waga (1993). In his paper, the time
varying cosmological term was defined by $\Lambda = \alpha R^{-2}
+ \beta H^{2} + \gamma$, where $\alpha$, $\beta$ and $\gamma$ are
arbitrary constants. Although rather different from the
deflationary ansatz assumed here, we see that in the limit $H <<
H_I$, our expression (5) reduces to $\rho_v \sim \beta\rho_T$, or
equivalently, $\Lambda \sim \beta (3H^{2} + 3kR^{-2})$. Therefore,
by considering only the flat case ($k=0$) and taking
$\alpha=\gamma = 0$, the late stages of the deflationary
cosmologies has exactly the same behavior of the models examined
by Waga. Note also that our $\beta$ parameter must be identified
with $1/3$ of that one appearing in his paper. More recently, some
aspects involving the angular size, luminosity distance, and SNe
type Ia observations in $\Lambda(t)$ models were also investigated
by Vishwakarma (2000, 2001). By generalizing earlier papers (Chen
and Wu 1990, Carvalho et al. 1992), He considered the interesting
ansatz $\Lambda = n\Omega H^{2}$, where $n$ is a dimensionless
constant and $\Omega$ is the density parameter of the fluid
component. Although proposed as a new cosmology, if we rewrite
this $\Lambda(t)$ in terms of the total energy density, this
decaying law corresponds precisely to the late stages of the
deflationary universes for $\beta = {n \over n + 3}$. A more
quantitative analysis accounting for angular sizes, the existence
of old high redshifts galaxies, and ages constraints from globular
clusters in this framework will be discussed in a forthcoming
communication. We also remark that models driven by a
gravitational adiabatic matter creation process (Calv\~ao et al.
1992, Lima and Alcaniz 1999, Alcaniz and Lima 1999) present some
kinematic expressions, like the dimensionless radial coordinate,
which are similar to the ones discussed here for $\Lambda(t)$
cosmologies (see Appendix). In these models, the present day
matter creation rate, $\psi_o = 3n_oH_o \approx 10^{-16}$ nucleons
$cm^{-3}yr^{-1}$, is also nearly the same rate transferred from
the decaying vacuum to the fluid component. However, these two
spacetimes are completely different from a physical viewpoint
because the negative creation pressure of the former scenario
cannot be interpreted as a genuine decaying vacuum component.

\begin{table} \label{Tabelabeta}
\begin{center}
\caption{Limits to $\beta$} 
\begin{tabular}{rlrll} \hline
\\
Test& \quad  \quad  \quad  \quad  & Open \quad  \quad  \quad   & Closed \quad  \quad  \quad   \\ 
\\
\hline
\\
Luminosity distance-redshift& & $\beta \le 0.83$ \quad  \quad  \quad & $\beta \le 0.58$\quad  \quad 
 \quad   \\
\\
Angular size-redshift& & $\beta \le 0.69$ \quad  \quad  \quad   & $\beta \le 0.56$\quad  \quad  
\quad   \\ 
\\
Number counts-redshift& & $\beta \le 0.19$ \quad  \quad  \quad   & $\beta \le 0.34$\quad  \quad  
\quad    \\ 
\\
\hline
\end{tabular}
\end{center}
\end{table}
\section{Conclusion}

The recent observational evidences for an accelerated state of the
present universe, obtained from distant SNe Ia (Perlmutter et al.
1999, Riess et al. 1998) gave a strong support to the search of
alternative cosmologies. As demonstrated here, a variable
$\Lambda$-term or a decaying vacuum energy density is also an
ingredient accounting for this unexpected observational result. In
the present paper we have analyzed all kinematic expressions for
flat, closed, and open cosmologies when the decaying vacuum is a
fraction of the total energy density. At the later stages of the
evolution, the rather slight changes introduced by $\Lambda(t)$,
which is quantified by the $\beta$ parameter, provides a
reasonable fit of several cosmological data. Kinematic tests like
luminosity distance, angular diameter and number-counts versus
redshift relations constrain perceptively the decaying vacuum (see
table 1). For models characterized by the pair ($\Omega_o,
\beta$), the age of the universe is always greater than the
corresponding FRW model ($\beta=0$), and even values bigger than
$H_o^{-1}$ are allowed for all values of the curvature parameter.

\appendix
\section{Dimensionless radial coordinate versus redshift relation}

Some observable quantities in the standard FRW model are easily
determined expressing the radial dimensionless coordinate $r$ of a
source light as a function of the redshift (Mattig 1958). In this
appendix, we derive a similar equation to the decaying vacuum
energy scenario discussed in this paper.

Now consider a typical galaxy located at $(r_1,\theta_1,\phi_1)$
emitting radiation to an observer at $(r=0,\theta_1,\phi_1)$. If
the waves leave the source at time $t_1$ and reach the observer at
time $t_0$, the null geodesic equation defining the light track
yields
\begin{equation}
\int_{t_0}^{t_1}
     \frac{dt}{R(t)} = \int_{0}^{r_1}
     \frac{dr}{\sqrt{1 - k{r}^{2}}}= \frac{sin^{-1}\sqrt
kr_1}{\sqrt k} = I\quad .
\end{equation}
Since $t=t(R)$, changing variable to $x={R \over R_o}$ and using
(8), the above result reads \small{
\begin{equation}
I = \frac{1}{R_o H_o} \int_{1 \over 1+z}^{1}{dx \over{x\sqrt{
1-\left(\frac{\Omega_o}{1-\beta}\right)
           + \left(\frac{\Omega_o}{1-\beta}\right)x^{-(1-3\beta)}}} }\quad.
\end{equation}}

This integral depends on the values of the $\Omega_o$ and $\beta$
parameters. For $\beta = \frac{1}{3}$ one finds the same results
of the coasting cosmology (Kolb 1989). For $\beta$ different of
${1 \over 3}$, we introduce a new auxiliary variable $y^{2} = [1 -
(\frac{1 - \beta}{\Omega_o})]x^{(1-3\beta)}$, in terms of which
the above equation becomes
\begin{equation}
\frac{sin^{-1}\sqrt kr_1}{\sqrt k} = \frac{\delta}{R_o
H_o(\frac{\Omega_o}{1-\beta} -
1)^{\frac{1}{2}}}\int_{\alpha_2}^{\alpha_1}
           \frac{dy}{\sqrt{1 - {y}^{2}}}\quad ,
\end{equation}
where $\delta = \frac{2}{(1-3\beta)}$, $\alpha_{1} = [1 - (\frac{1
- \beta}{\Omega_o})]^{\frac{1}{2}}$, and $\alpha_2 =
\alpha_{1}(1+z)^{-\frac{(1-3\beta)}{2}}$.

The right hand side of the above integral is the same appearing in
(A.1) for $k=1$. Hence, replacing in (A.3) the value of $k$ given
by (\ref{rho}) and (7), it is readily seen that
\begin{equation}
r_1(z) = \frac{sin[\delta sin^{-1}\alpha_1- \delta
sin^{-1}\alpha_2]}{R_oH_o(\frac{\Omega_o}{1-\beta} -
1)^{\frac{1}{2}}}\quad .
\end{equation}
In particular, the limit for a flat universe ($\Omega_o=1$) yields
\begin{equation}
r_1(z)= {\frac{2}{(1-3\beta)R_oH_o}\{1 - (1 + z)^{\frac{2}{{1
-3\beta}}}\}}\quad ,
\end{equation}
which could have been obtained directly from (A.2).

In terms of the deceleration parameter (A.4) may be rewritten as
\begin{equation}
r_1(z) = \frac{sin[\delta sin^{-1}\alpha_1 - \delta
sin^{-1}\alpha_2]}{R_oH_o(\frac{2q_o}{1 - 3\beta} -
1)^{\frac{1}{2}}}\quad ,
\end{equation}
which in the limit $\beta \rightarrow 0$ reduces to the usual FRW
result (Weinberg 1972)
\begin{equation}
r_1(z) = \frac{q_oz + (q_o - 1)(\sqrt{2q_oz + 1} -
1)}{H_oR_oq_o^{2}(1 + z)}\quad .
\end{equation}
Equation (A.4), or equivalently (A.6), plays a key role in the
derivation of some astrophysical quantities discussed in this
paper.

{\bf Acknowledgements:} The authors are grateful to J. S. Alcaniz
for helpful discussions and a critical reading of the manuscript.
This work was partially supported by the project Pronex/FINEP (No.
41.96.0908.00), and CNPq (62.0053/01-1-PADCT III/Milenio).

\end{document}